\documentclass{appolb}
\usepackage{epsfig}
\newcommand{\be}{\begin{eqnarray}}
\newcommand{\ee}{\end{eqnarray}}
\newcommand{\beq}{\begin{equation}}
\newcommand{\eeq}{\end{equation}}
\newcommand{\nn}{\nonumber}
\newcommand{\qq}{\langle \bar{q}q\rangle}
\newcommand{\la}{\lambda}

\begin{document}
\title{Equivalence of Matrix Models for Complex QCD Dirac Spectra%
\thanks{Presented at the Workshop on Random Geometry in Krakow 
May 2003, Poland}%
}
\author{G. Akemann
\address{Service de Physique Th\'eorique CEA/Saclay\\
Unit\'e associ\'ee CNRS/SPM/URA 2306\\
F-91191 Gif-sur-Yvette Cedex, France
}
%\and
%the Name(s) of other Author(s)
%\address{and their affiliation}
}
\maketitle
\begin{abstract}
Two different matrix models for QCD with a non-vanishing quark chemical 
potential are shown to be equivalent by mapping the corresponding partition 
functions. The equivalence holds in the phase with broken chiral symmetry. 
It is exact in the limit of weak non-Hermiticity, where the chemical potential
squared is rescaled with the volume. At strong non-Hermiticity it holds only 
for small chemical potential. The first model proposed by Stephanov is 
directly related to QCD and allows to analyze the QCD phase diagram. In the 
second model suggested by the author all microscopic spectral correlation 
functions of complex Dirac operators can be calculated in the broken phase. 
We briefly compare those predictions to complex Dirac eigenvalues from 
quenched QCD lattice simulations.
\end{abstract}
  
\section{Introduction}

Ten years ago the proposal \cite{SV} was made to 
use random matrix models in the study of the low energy phase 
of Quantum Chromodynamics (QCD), and it has turned out to be very successful. 
Since then many new matrix model correlation functions have been calculated 
for that purpose. 
The field theoretic origin for the applicability of matrix models 
has been understood and 
a wealth of data from QCD lattice simulations has been compared to their 
predictions. We refer the reader to \cite{JT} for a review summarizing these 
developments.
However, the introduction of a chemical potential $\mu$ for quarks 
has left many open questions until very recently, 
concerning both numerical simulations as well as analytical matrix model 
predictions. 
The reason is that for non-vanishing $\mu$ the eigenvalues and thus the 
determinant of the Dirac operator become complex.

Let us summarize what can be calculated using random matrix models for QCD. 
The predictions can be put in two groups: the Dirac operator spectrum 
and the phase diagram. 
In the first case the distribution of Dirac eigenvalues $\la_i$
rescaled with the volume $V$, $V\la_i=const.$, are predicted from matrix models
close to the origin $\la=0$. This region is magnified microscopically
as it is most sensitive to spontaneous chiral symmetry breaking.
This fact can be seen from the Banks-Casher relation, 
$\rho(0)=V\qq/\pi$, which relates the Dirac operator spectral density 
$\rho(\la)$ at the origin to the order parameter for chiral symmetry breaking
$\qq$, the chiral condensate. 
The predictions of matrix models are universal and 
depend only the $N_f$ quark flavors, their masses $m_f$, 
the topological charge $\nu$ and the global symmetry breaking pattern. 
They can be related to a particular limit 
of chiral perturbation theory, in which only the zero momentum modes 
of the Goldstone bosons contribute. The matrix model description breaks down 
at the Thouless energy, when non-zero momentum modes come into play.

In the second application matrix models are used as an effective model 
for QCD to predict the qualitative features of the 
phase diagram. The order and location of the phase 
transition lines is found by inspecting the non-analyticities of the matrix 
model partition function in terms of parameters as the temperature $T$ and 
chemical potential $\mu$. 
Here, the close relation to field theory is lost
and the matrix model resembles a Landau-Ginsburg description. 
The following result was obtained in \cite{Halasz1} 
for QCD with two massless quark flavors.
\begin{figure}[h]
\unitlength1cm
\begin{picture}(12.2,5.5)
\put(2.0,0.3){\vector(1,0){8}}
\put(10.2,0.2){$\mu$}
\put(2.0,0.3){\vector(0,1){5}}
\put(1.1,5.1){$T$}
\qbezier[30](2.0,4.0)(4.0,3.7)(5.0,3.5)
\put(5.0,3.4){$\bullet$}
\qbezier(5.0,3.5)(7.5,3.0)(9.0,0.3)
\put(3.1,1.5){$\qq\neq0$}
\put(8.1,3.8){$\qq=0$}
\end{picture}
\label{phase} 
\caption{The schematic QCD phase diagram with $N_f=2$ massless flavors from 
\cite{Halasz1}.}
\end{figure}
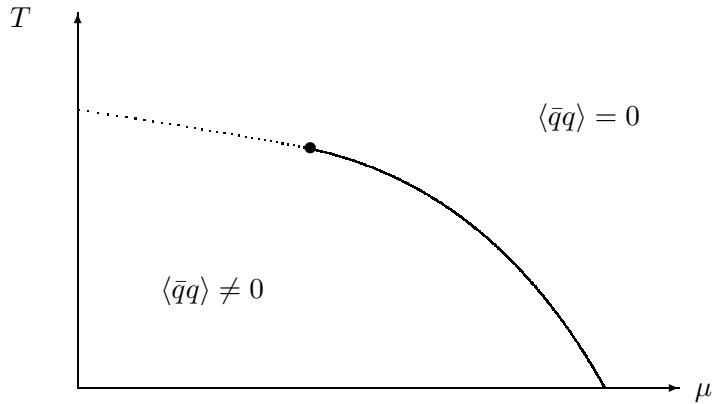
A first order (full) line starting at large $\mu$ merges with a second order 
(dotted) line from $\mu=0$ into a tri-critical point. 
Other transitions from  nuclear matter formation or more 
complicated phases such as color super-conductivity have been omitted here, 
where the latter can be addressed using matrix models as well \cite{Benoit}.

We take this simplified phase diagram to explain which regions 
are explored. 
One part of the phase diagram accessible to QCD lattice simulations 
for three colors 
is the temperature axis at $\mu=0$ (including more flavors and 
the influence of their mass ratio). The second part is for small values of 
$\mu$ in the vicinity of  
the second order phase transition line up to or crossing the 
tri-critical point. Only in the quenched approximation the full diagram can 
be covered so far. We refer to \cite{Kogut} for reviews and 
references.
On the other hand very detailed matrix model predictions 
for the microscopic Dirac spectrum are available at $T=\mu=0$ (see 
\cite{JT} and references therein). 
For $T\neq0$ and $\mu=0$ analytic predictions exist too \cite{TT,Romuald}
and have been compared to lattice 
simulations along the $T$-axis up to and above the transition 
\cite{Poul&Kim}. 

Here, we will investigate the $\mu$-axis at $T=0$ instead, 
for which two different matrix models have been suggested 
\cite{Steph,A02}. The model introduced first by Stephanov \cite{Steph} 
explicitly includes a chemical potential, using the same symmetry arguments 
as \cite{SV}. However, in this model only global properties are available,  
with the microscopic spectrum being unknown. For that reason 
another model \cite{A02} was proposed where all microscopic spectral 
correlators can 
be calculated. This model is only valid in the broken phase by construction.
It follows from \cite{SV} by analytic continuation of the corresponding 
orthogonal polynomials into the complex plane and contains 
$\mu$ only implicitly through a non-Hermiticity parameter $\tau$. 
Our aim here is to show that \cite{A02} is equivalent to \cite{Steph} and 
thus as closely related to QCD.
We prove that the two partition functions of these models 
are equal in the broken phase in the limit of infinite matrix size. 
In the limit of weak non-Hermiticity, keeping $V\mu^2$ fixed, 
the equivalence is shown for $N_f=1,2$ and $3$ mass-degenerate flavors 
(the case of general $N_f$ and $\nu\neq0$ 
will be treated elsewhere \cite{AFV}). 
At strong non-Hermiticity the partition functions are mapped for $N_f=1$
to leading order in small $\mu$, and we expect the equivalence 
to hold also for $N_f\geq2$. 
The equality of partition functions 
does not automatically imply that all 
correlation functions are identical as well. 
However, it makes it very plausible that the results for the 
microscopic spectrum \cite{A02} are common to both models and 
are universal.
We also give a brief comparison between the predictions \cite{A02} 
and quenched QCD lattice simulations \cite{AW,AWII} for small 
$\mu\neq0$ in the broken phase.

%%%%%%%%%%%%%%%%%%%%%%%%%%%%%%%%%%%%%%%%%%%%%%%%%%%%%%%%%%%%%%%%%%%%%%%%

\section{Equivalence of matrix models for QCD with $\mu\neq0$
}

The following matrix model was introduced by Stephanov \cite{Steph} to study 
QCD with chemical potential 
\beq
{\cal Z}_I(\mu;\{m_f\}) \equiv \int d\Phi
\prod^{ N_f}_{f=1} 
\det\left(
\begin{array}{c}
m_f      \ \ \ \             i\Phi + \mu\\
\\
i\Phi^\dagger + \mu  \ \ \ \     m_f\\
\end{array}
\right) \exp[-N\qq^2\Tr\Phi^\dagger\Phi]
\label{Z1}
\eeq
Here $\Phi$ is a complex matrix of size $N\times(N+\nu)$ with the partition 
function being in the sector of topological charge $\nu$.
The quark masses $m_f$ of the $N_f$ flavors have to be rescaled with the 
volume in the same way as the eigenvalues in the microscopic limit.
The block structure of the Dirac matrix 
follows from chiral symmetry and the location of $\mu$ is dictated 
by adding $\mu q^\dagger q$ to the quark Lagrangian.
The Gaussian weight is chosen such that the Banks-Casher relation for $\mu=0$
is satisfied. In the following we will mostly set $\qq=1$ for simplicity.

The partition function for $N_f=1$ quark flavor was already know 
\cite{Halasz2}, 
\beq
{\cal Z}_I(\mu; m)\ =\ \exp[-N(1+\mu^2)]\exp[-Nm^2]\ 
I_0\left(2Nm\sqrt{1+\mu^2}\right)\ ,
\label{Z1N}
\eeq
where $I_0$ denotes the modified Bessel function. Here we have taken already 
the large-$N$ limit, assuming that the mass has to be rescaled 
with an appropriate power of the volume $V=2N$ (see eqs. (\ref{micro}), 
(\ref{microstrong})). 
Eq. (\ref{Z1N}) holds in the broken phase. At a 
critical chemical potential $\mu_c\approx 0.527$ 
determined by the equation \cite{Halasz2}
\beq
1+\mu_c^2 +\log\mu_c^2 \ =\ 0\ , 
\label{mucrit}
\eeq
a first order phase transition takes place. Above $\mu>\mu_c$ it holds
${\cal Z}_I(\mu; m)\sim\mu^{2N}$ for $m=0$ 
\cite{Halasz2}.
This can be show by mapping eq. (\ref{Z1}) to a sigma model partition 
function over complex matrices $\sigma$ of size $N_f\times N_f$ in
eq. (\ref{Zsigma}), allowing us to treat also $N_f>1$ in the broken 
phase below. 
The macroscopic spectral density of complex eigenvalues $z=x+iy$ given by
\cite{Steph} 
\beq
\rho(z;\mu)= \frac{1}{4\pi}\left(\frac{\mu^2+y^2}{(\mu^2-y^2)^2} -1\right), 
\label{rhomacro1}
\eeq
is bounded by the following curve
\beq
0= y^4(x^2+4\mu^2)\ +\ y^2(1+2\mu^2(2-x^2-4\mu^2))\ +\ \mu^4(x^2-4(1-\mu^2))\ .
\label{bound}
\eeq
It splits into two domains only at $\mu=1>\mu_c$.
On the other hand for small $\mu\ll1$ the density 
simplifies to a constant on an ellipse 
\beq
\rho(z;\mu)= \frac{1}{4\pi\mu^2}\ , \ \ 
\ \mbox{if} \ \ \ \ \frac{x^2}{4}+\frac{y^2}{4\mu^4}\leq 1 \ .
\label{rhomu}  
\eeq

Let us compare this to the second model \cite{A02}.  
The quenched $N_f=0$ partition function is defined as
\be
{\cal Z}_{II}(\tau) &\equiv&  \int \prod_{j=1}^N
\left(d^2z_j\ w(z_j,z_j^\ast) \right)
\left|\Delta_N(z^2)\right|^2, 
\label{Z2q}\\
w(z,z^\ast)&=& |z|^{2\nu+1}\! \exp\left[
-\frac{N}{1-\tau^2}\left(|z|^2 -\frac{\tau}{2}(z^2+z^{2\,\ast})\right)
\right]\ ,
\label{weight}
\ee
given in terms of complex eigenvalues instead of matrices. 
The Vandermonde determinant is denoted by 
$\Delta_N(z^2)=\prod_{k>l}^N(z_k^2-z_l^2)$.
The parameter $\tau\in[0,1]$ labels the degree of non-Hermiticity 
interpolating between real and maximally complex eigenvalues for $\tau=1$ and
$\tau=0$, respectively. 
The macroscopic density reads\footnote{Note the missing factor 2 in the 
corresponding eqs. in \cite{A02}.} 
\beq
\rho(z;\tau)= \frac{1}{2\pi(1-\tau^2)}\ , \ \ 
\ \mbox{if} \ \ \ \ \frac{x^2}{2(1+\tau)^2}+\frac{y^2}{2(1-\tau)^2}\leq 1 \ .
\label{rhotau} 
\eeq
Identifying the two macroscopic densities eqs. (\ref{rhomu}) and 
(\ref{rhotau}), which implies that to leading order
in $N$ all moments of the two models become equal, leads us to$^1$
\beq
2\mu^2 \ \equiv\ (1-\tau^2) \ .
\label{mutau}
\eeq
Clearly the model eq. (\ref{Z2q}) is always in the broken phase 
as the density is constant on a single domain for all $\tau\in[0,1)$. 
Here, all microscopic $k$-point correlations functions can be explicitly 
calculated using the technique of orthogonal polynomials in the complex plane
\cite{A02} (see e.g. eqs. (\ref{weakrho}), (\ref{strongrho}) below).
For that purpose we define the following complex Laguerre polynomials
in monic normalization,
\be
\tilde{P}_k^{(\nu)}(z)\ \equiv\ 
(-1)^k k!\left(\frac{2\tau}{N}\right)^k 
L_k^{(\nu)}\left(\frac{Nz^2}{2\tau}\right) \ , 
\label{Laguerre}
\ee
which are orthogonal with respect to the weight eq. (\ref{weight}).

In order to compare to the partition function eq. (\ref{Z1}) above we also 
have to introduce mass terms here, which we do as follows:
\be
 {\cal Z}_{II}(\tau;\{m_f\}) &\equiv&  {{\cal Z}_{II}(\tau)}^{-1}
\int \prod_{j=1}^N
\left(d^2z_j\ w(z_j,z_j^\ast)\prod_{f=1}^{N_f}(z_j^2+m_f^2)\ 
 \right)
\left|\Delta_N(z^2)\right|^2 \nn\\
&=&
\left\langle  \prod_{j=1}^N \prod_{f=1}^{N_f}(z_j^2+m_f^2)\right\rangle \ .
\label{Z2}
\ee
Such an expectation value can be calculated using the theorem proved in 
\cite{AVII} for characteristic polynomials. In our case it only contains 
orthogonal polynomials in monic normalization, and we obtain
\beq 
{\cal Z}_{II}(\tau;\{m_f\}) \sim \frac{\det_{k,l=1,\ldots,N_f}\left[
\tau^{N+k-1} L_{N+k-1}^{(\nu)}\left(\frac{-Nm_l^2}{2\tau}\right) 
\right]}{\Delta_{N_f}(m^2)}\ .
\label{Z2N}
\eeq
Here, we only display the $\tau$-dependent prefactors.
The partition function defined in eq. (\ref{Z2})
is thus real as it should. From eq. (\ref{Z2N}) we can take the 
large-$N$ limit where one has to distinguish now two cases, weak and strong
non-Hermiticity. 

%%%%%%%%%%%%%%%%%%%%%%%%%%%%%%%%%%%%%%%%%%%%%%%%%%%%%%%%%%%%%%%%%%%%%
\subsection{The weak non-Hermiticity limit}

In the weak non-Hermiticity limit introduced by \cite{FKS} 
(see also the recent review \cite{FS}) we take the Hermitian limit
$\tau\to1$ and the infinite volume limit $N\to\infty$ at the same time, 
keeping 
\beq
N(1-\tau^2)\ =\ 2N\mu^2 \ \equiv \ \alpha^2
\label{weaklim}
\eeq
finite. In this limit the macroscopic density eq. (\ref{rhotau})
becomes localized on the real line while the microscopic correlations 
functions still extend into the complex plane. 
In the microscopic limit we have to rescale the eigenvalues and 
consequently also the masses with the volume. In order to obtain a parameter 
free prediction this is done by 
\beq
2N\qq z\ =\ \xi\ ,
\label{micro}
\eeq
and similarly for the masses.
Since in the limit $\tau\to1$ the model eq. (\ref{Z2q}) has $\qq=1/\sqrt{2}$
we define the rescaled masses 
as $\sqrt{2}Nm_f=\omega_f$.
Taking the asymptotic large-$N$ limit 
of eq. (\ref{Z2N}) in this way we obtain
\beq
{\cal Z}_{II, weak}(\alpha;\{\omega_f\})
\sim \exp\left[-\frac{N_f}{2}\alpha^2\right] 
\frac{\det_{k,l=1,\ldots,N_f}\left[\omega^{k-1}_l I_{k-1+\nu}(\omega_{l})
\right]}{\Delta_{N_f}(\omega^2)}\ ,
\label{Zweak}
\eeq
where we have used 
that $\tau^N\to\exp[-\alpha^2/2]$ from the weak limit eq. (\ref{weaklim}). 
The expression eq. (\ref{Zweak}) only differs from the finite volume 
partition function at $\mu=0$ \cite{LS}
by the $\alpha$-dependent exponential prefactor.

In order to prove that the same results can be obtained from the 
matrix model partition function eq. (\ref{Z1}) we rewrite it in terms 
of the following $\sigma$-model representation \cite{Halasz2}:
\be
{\cal Z}_I(\mu;m) &=& \int d\sigma\ \exp[-N\qq^2\Tr\sigma\sigma^\dagger]
\ \det\left(
\begin{array}{cc}
\sigma+m & \mu\\
& \\
\mu      & \sigma^\dagger + m\\
\end{array}
\right)^N 
\label{Zsigma}
\ee
Here, we have restricted ourselves to 
mass-degenerate flavors and to
the sector of topological charge $\nu=0$. Compared to eq. (\ref{Z1}) 
$N$ and $N_f$ have interchanged r\^{o}les, as 
$\sigma$ is now a complex $N_f\times N_f$ matrix. 
We show explicitly for $N_f=2$ 
how to calculate the integral eq. (\ref{Zsigma}) by parameterizing $\sigma$.
The calculation for $N_f=3$ then follows in a similar way.
Any complex matrix $\sigma$ can be written in the Schur decomposition 
\beq
\sigma \ =\ U (Z+R) U^{-1}\ ,
\eeq
where $U$ is unitary, $Z=$diag$(z_1,\ldots,z_{N_f})$ 
is a diagonal matrix containing the complex 
eigenvalues and $R$ is a complex, 
strictly upper triangular matrix. For $N_f=2$ we thus 
have two complex eigenvalues and $R$ contains a single complex number. Setting 
$\qq=1$ we can write 
\be
{\cal Z}_I(\mu;m) &=&  \int d^2z_1\,d^2z_2\,d^2R
\ \exp\left[-N(|z_1|^2+|z_2|^2+|R|^2)\right]\ |z_2-z_1|^2\nn\\
&&\times\det\left(
\begin{array}{c}
|z_1|^2+m(z_1+z_1^*)+m^2-\mu^2 \ \ \ \ \ \ \ \  (z_1^*+m)R\\
 \\
R^*(z_1+m)              \ \ \ \ \ \ \ \  |z_2|^2+m(z_2+z_2^*)+m^2-\mu^2\\
\end{array}
\right)^N\nn\\
&=& \int_0^\infty\!\!dr_1r_1dr_2r_2d\rho\rho \int_{-\pi}^{\pi}
\!d\varphi_1d\varphi_2d\phi\,
[(r_1^2-\mu^2)(r_2^2-\mu^2)-\rho^2\mu^2]^N\nn\\
&&\times\exp\left[-N(\Sigma_{j=1,2}(r_j^2-2r_jm\cos\varphi_j)+2m^2+\rho^2)
\right]\nn\\
&&\times\left( r_1^2+r_2^2
-2r_1r_2(\cos\varphi_1\cos\varphi_2+\sin\varphi_1\sin\varphi_2)\right)\nn\\
&=&
(2\pi)^3\int_0^\infty dr_1r_1dr_2r_2d\rho\rho\ 
[(r_1^2-\mu^2)(r_2^2-\mu^2)-\rho^2\mu^2]^N\nn\\
&&\times\exp\left[-N(r_1^2+r_2^2+\rho^2\!+m^2)\right]\ 
[ (r_1^2+r_2^2)I_0(2Nmr_1)\nn\\
&&\times I_0(2Nmr_2)
-2r_1r_2I_1(2Nmr_1)I_1(2Nmr_2)]\ .
\label{Zexact}
\ee
In the first step we have applied the Schur decomposition, which leads to the 
Vandermonde determinant. In the second step we have changed variables, 
$z_j\to \tilde{z}_j=(x_j+m)+i\,y_j=r_j\exp[i\varphi_j]$ for $j=1,2$, 
introducing polar coordinates here and in 
$R=\rho\exp[i\phi]$. In the last step we have carried out the angular 
integrations. So far all steps are exact. 
Next we employ the scaling eq. (\ref{weaklim}), $\mu^2=\alpha^2/(2N)$. 
Consequently we can replace the first factor in the last equation above by an 
exponential 
\beq
[(r_1^2-\mu^2)(r_2^2-\mu^2)-\rho^2\mu^2]^N\to
(r_1r_2)^N\exp[-\alpha^2(\rho^2+r_1^2+r_2^2)/(2r_1^2r_2^2)]\ ,
\eeq
using $\lim_{N\to\infty}(1-x/N)^N=\exp[-x]$. 
The integral over $\rho$ thus reads:
\be
&&\lim_{N\to\infty} \int_0^\infty d\rho\rho\ \exp[-N\rho^2]\ 
[(r_1^2-\mu^2)(r_2^2-\mu^2)-\rho^2\mu^2]^N\nn\\
&&= \frac{(r_1^2r_2^2)^{N+1}}{2Nr_1^2r_2^2+\alpha^2}
\exp\left[\frac{-\alpha^2(r_1^2+r_2^2)}{2r_1^2r_2^2}\right].
\ee
If we neglect the subleading $\alpha^2$ 
in the denominator on the right hand side 
the remaining integrals over $r_1$ and $r_2$ in eq. (\ref{Zexact})
completely factorize. We can thus investigate them separately 
using a saddle point approximation. 
Because of $\qq=1$ we rescale the mass here as $2Nm=\omega$. 
Neglecting higher orders in $1/N$ the saddle point is at $r_{j=1,2}^2=1$. 
The two 
contributions containing the factors  
$I_0$ and $I_1$ can be written in the following form, 
\be
{\cal Z}_{I,weak}(\alpha;\omega)\ \sim\ 
\exp\left[-\alpha^2\right]
\left|
\begin{array}{cc}
I_0(\omega) & I_1(\omega) \\
& \\
I_1(\omega) & I_0(\omega)  \\
\end{array}
\right|\ .
\label{Zweakdeg}
\ee
This is equivalent to eq. (\ref{Zweak}) for $N_f=2$, 
when taking the limit of degenerate masses there. In particular it 
justifies on the level of partition functions 
the identification eq. (\ref{mutau}), leading to the scaling 
eq. (\ref{weaklim}). 
The same calculation can be easily done for $N_f=3$, following the same lines 
as above. In fact, recently we managed to show in full
generality, that eq. (\ref{Zsigma}) leads to 
eq. (\ref{Zweak}) for all $N_f$ with non-degenerate masses and 
topological charge $\nu\neq0$ \cite{AFV}.

%%%%%%%%%%%%%%%%%%%%%%%%%%%%%%%%%%%%%%%%%%%%%%%%%%%%%%%%%%%%%%%%%%%%%
\subsection{The strong non-Hermiticity limit}

In this limit $\tau\in[0,1)$ and thus $\mu^2$ are kept fixed while the 
large-$N$ limit is taken. Therefore both macroscopic and microscopic 
correlators extend to the complex plane. The microscopic rescaling 
has to be modified, reading
\beq
\sqrt{2N}\qq z\ =\ \xi\ 
\label{microstrong}
\eeq 
for the eigenvalues and masses. 
Because $\mu^2$ does not scale with $N$, the saddle point analysis 
for $N_f=2$ or more flavors is difficult, as can be seen from eq. 
(\ref{Zexact}). We therefore restrict ourselves to the case $N_f=1$, taking 
eq. (\ref{Z1N}) as a starting point. If we rescale the mass according to 
eq. (\ref{microstrong}), $\sqrt{2N}m=\omega$, and expand for small values 
of $\mu^2\ll1$, we obtain
\beq
{\cal Z}_{I,strong}(\mu; \omega)= \exp[-N(1+\mu^2)]
\exp\left[\frac{-\omega^2}{2}\right] I_0\left(\sqrt{2N}\omega(1+\frac12
\mu^2)\right).
\label{ZIstr}
\eeq
The evaluation of the second partition function takes a little more care, 
starting from eq. (\ref{Z2N}). Expressing the single Laguerre polynomial 
as an integral, we obtain
\be
{\cal Z}_{II}(\tau; m) &\sim& \tau^N\exp\left[\frac{-Nm^2}{2\tau}\right]
\int_0^\infty ds \exp[-Ns] s^N I_0\left(Nm\sqrt{\frac{2s}{\tau}}\right)\nn\\
&\sim& \mbox{e}^{-N} \tau^N \exp\left[\frac{-Nm^2}{2\tau}\right]
 I_0\left(Nm\sqrt{\frac{2}{\tau}}\right) \ .
%\label{}
\ee
In the saddle point expansion is the second step we have already used the 
mass rescaling $\sqrt{N}m=\omega$, which is appropriate here for 
$\qq^2=1/2$ with $\tau$ close to unity. If we employ again the relation
(\ref{mutau}), $\tau^2=1-2\mu^2$, 
insert it here and expand in $\mu^2\ll1$ we arrive at
\be
{\cal Z}_{II,strong}(\mu; \omega)&=& \exp[-N(1+\mu^2)]
\exp\left[\frac{-\omega^2}{2}(1+\mu^2)\right] \nn\\
&&\times I_0\left(\sqrt{2N}\omega(1+\frac12\mu^2)\right)
\label{ZIIstr}
\ee
We have used that $\tau^N\to\exp[-N\mu^2]$, omitting terms of order 
$O(\mu^4)$. We observe that the leading order (suppression) term in 
$\mu^2$ as well as the term $I_0$ exponentially growing in $\omega$ match 
to the given order $O(\mu^2)$. We recall that the 
matching of the two macroscopic densities 
also included an expansion of eq. (\ref{rhomacro1}) in $\mu^2$ 
to obtain eq. (\ref{rhomu}).
The finite term $\exp[-\omega^2/2]$ only matches to zero-th order. 

We find this mapping of the two partitions functions at strong non-Hermiticity
quite remarkable and can show it to hold also for $N_f\geq2$ and small 
$\mu^2\ll1$ \cite{AFV}. Another reason for that 
is that the strong limit can be obtained from the weak limit in taking 
$\alpha\to\infty$. 
While the two partition functions may disagree for larger values of $\mu^2$, 
as the expansions indicates, this does not exclude a universal behavior of 
the corresponding microscopic correlation functions. A mapping of these 
quantities may though include a nontrivial, $\mu$- and $\tau$-dependent 
unfolding procedure, respectively.

%%%%%%%%%%%%%%%%%%%%%%%%%%%%%%%%%%%%%%%%%%%%%%%%%%%%%%%%%%%%%%%%%%%%%%%%
\section{Comparison to lattice data}\label{data}

In this section we briefly recall the matrix model predictions 
\cite{A02} and their confirmation 
by quenched QCD lattice data \cite{AW}. A more detailed analysis of the data 
will be published elsewhere \cite{AWII}. 

The quenched microscopic spectral density at weak non-Hermiticity 
in the sector of topological charge $\nu$ is given by 
\beq
\rho_{weak}(\xi) = \frac{\sqrt{\pi\alpha^2}}{\mbox{erf}(\alpha)}\ |\xi|\ 
\exp\!\left[\frac{-1}{\alpha^2}(\Im m\,\xi)^2\right]
\int_0^1 dt\ \mbox{e}^{-\alpha^2t} 
J_\nu(\sqrt{t}\xi)J_\nu(\sqrt{t}\xi^\ast)\:,
\label{weakrho}
\eeq
where the eigenvalues are rescaled according to eq. (\ref{micro}). 
The density is normalized to unity for large arguments on the real line. 
At strong non-Hermiticity we obtain 
\beq
\rho_{strong}(\xi) = \sqrt{\frac{2\pi}{1-\tau^2}}
\ |\xi|\ \exp\!\left[\frac{-|\xi|^2}{1-\tau^2}\right]
I_\nu\left(\frac{|\xi|^2}{1-\tau^2}\right),
\label{strongrho}
\eeq
with the different rescaling eq. (\ref{microstrong}). Higher order correlation 
functions can be found in \cite{A02}. 

The data we compare with were generated on a $6^4$ lattice at gauge coupling 
$\beta=5.0$. Since staggered fermions at strong coupling away from the 
continuum limit are topology blind as explained in \cite{Poulnu}, 
%and they show a strong mixing 
%between zero and non-zero eigenvalues 
we have set $\nu=0$ in all comparisons 
below. The two values chosen for the chemical 
potential are $\mu=0.006$ and $\mu=0.2$. They correspond to weak and strong 
non-Hermiticity, respectively. We display the same figures as in ref. \cite{AW}
but with more statistics (17.000 and 20.000 configurations, respectively). 
For simulations with other lattice sizes, confirming 
the scaling prediction eq. (\ref{weaklim}) in the weak limit, 
we refer again to  \cite{AWII}.
\begin{figure}[-h]
\centerline{
\epsfig{figure=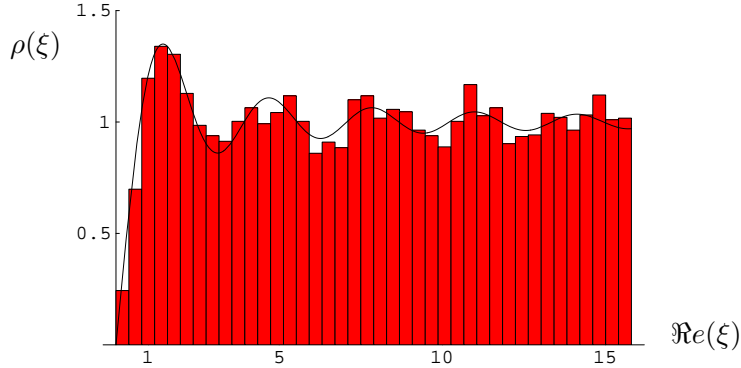,width=18pc}
\put(10,10){$\Re e(\xi)$}
\put(-240,120){$\rho(\xi)$}
}
\caption{
  Cut of the Dirac eigenvalue density along the real axis for
  $\mu=0.006$.
\label{Recut}}
\end{figure}
\begin{figure}[-h]
\centerline{
\epsfig{figure=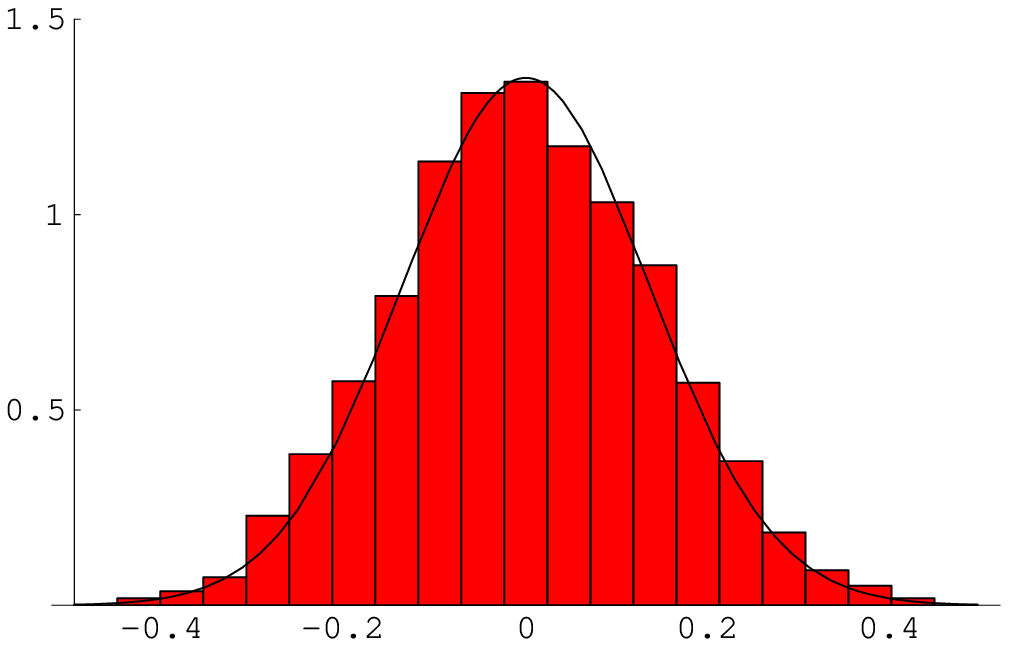,width=15pc}
\epsfig{figure=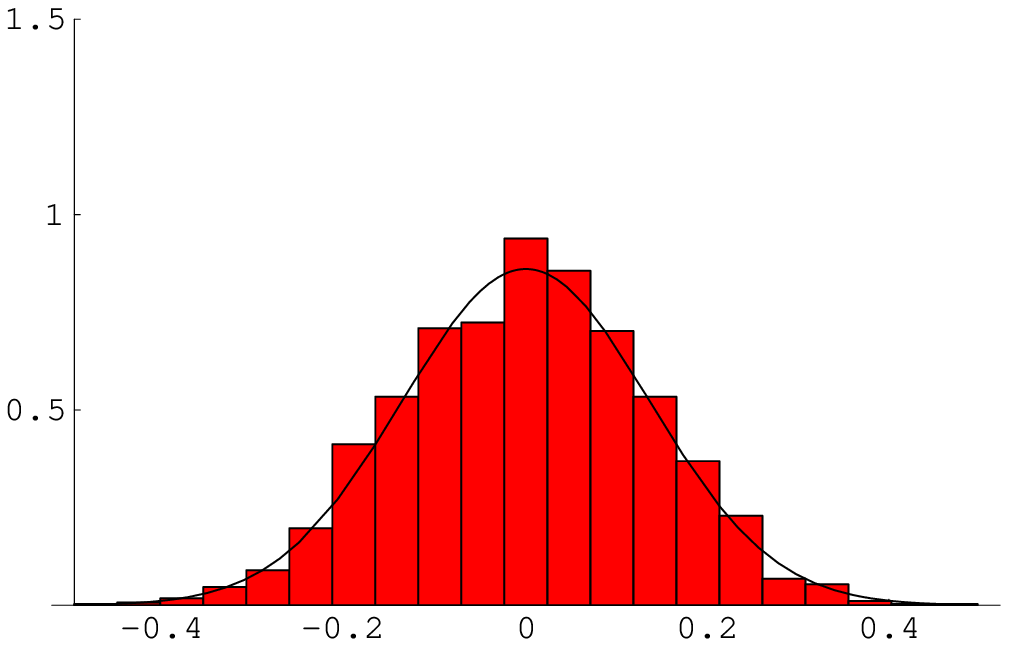,width=15pc}
\put(-30,20){$\Im m(\xi)$}
\put(-375,95){$\rho(\xi)$}
\put(-210,20){$\Im m(\xi)$}
\put(-192,95){$\rho(\xi)$}
}
\caption{
  Cuts of the same density along the imaginary axis, 
  with fixed real part of the eigenvalues at the first
  maximum (left) and minimum (right) in fig.~\ref{Recut}.
\label{Imcut}}
\end{figure}
The data are rescaled with the mean level spacing $d\sim 1/V$
determined as follows.
For $\mu=0.006$ the data almost lie on the real axis and are therefore 
ordered according to their real part. We thus simply count the number 
of eigenvalues in a (real) interval of given length to determine the 
average spacing. The same factor $d$ is used to rescale $\mu^2$
from eq. (\ref{weaklim}) and we obtain $\alpha=0.19$. In fig. \ref{Recut} 
and \ref{Imcut} 
eq. (\ref{weakrho}) is plotted with this value of $\alpha$ versus the rescaled 
data, and we find an excellent agreement. We note that there are no free 
parameters and that no fit has been made\footnote{The 
correction of eq. (\ref{mutau}) explains the small 
discrepancy found in \cite{AW}.}.
Fig. (\ref{Recut}) is very reminiscent to comparisons for $\mu=0$ (see 
\cite{JT}), with the difference that we need at least 10 times more statistics.
This is due to the spread of the data into the complex plane
seen in fig. \ref{Imcut}, which is thus truly testing the value of 
$\alpha$.
Here, eq. (\ref{weakrho}) is plotted for fixed real part $\Re e(\xi)$ 
on the first maximum and minimum (saddle) next to the origin. Comparing to the 
histograms that contain the maximum or minimum we find an excellent 
agreement. In \cite{AWII} the same picture is confirmed 
for different lattice sizes keeping the value of $\alpha$ fixed.

\begin{figure}[h]
%\epsfxsize=10cm   %width of figure - will enlarge/reduce the figures
%\epsfbox{fig3.eps}
%\figurebox{2cm}{3cm}{} %to have a box alone 
\vspace*{-2mm}
\centerline{\epsfxsize=2.22in\epsfbox{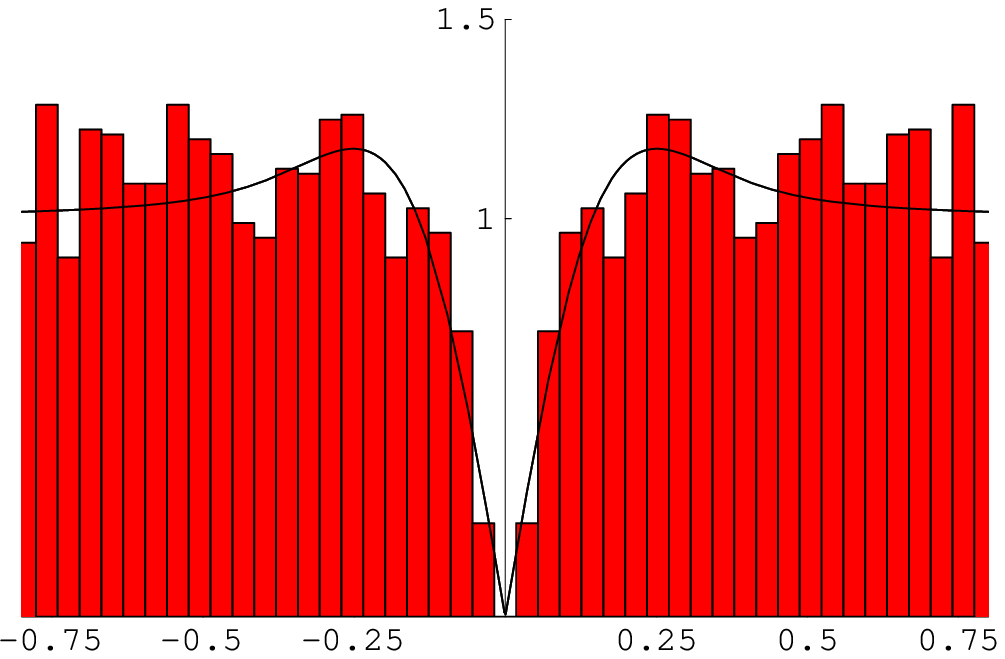}
\epsfxsize=2.22in\epsfbox{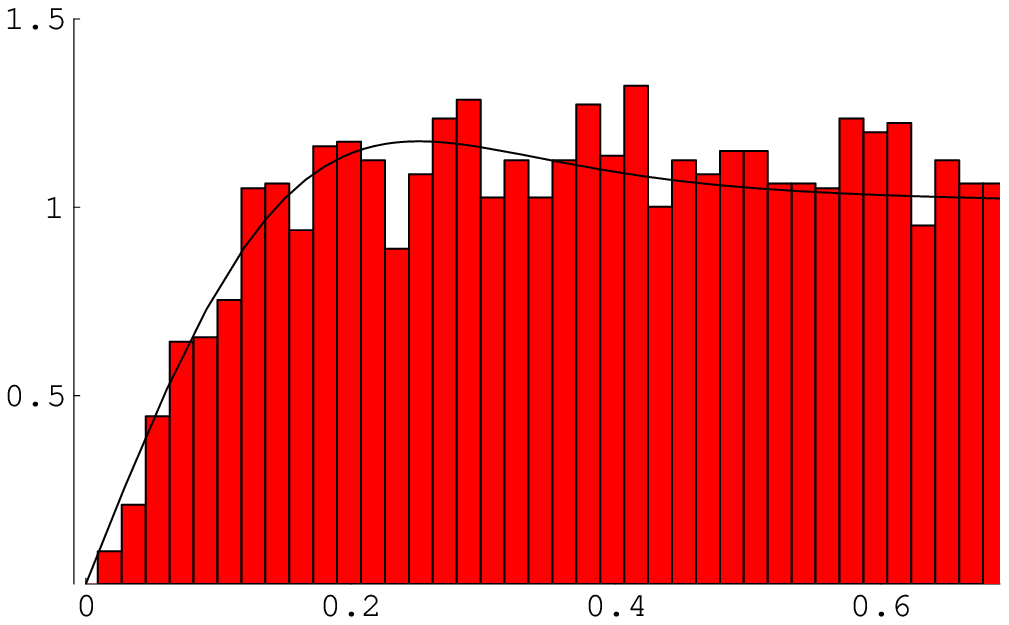}
\put(2,10){$\Re e(\xi)$}
\put(-275,98){$\rho(\xi)$}
\put(-352,10){$\Im m(\xi)$}
\put(-145,96){$\rho(\xi)$}
}   
\caption{
Cuts of the Dirac eigenvalue density for $\mu=0.2$ 
along the imaginary (left) and real axis (right).
\label{2Dcutstr}}
\end{figure}
At strong non-Hermiticity $\mu$ does not scale with the volume and we thus 
have $\tau=\sqrt{1-2\mu^2}\approx0.96$  for $\mu=0.2$. The data are rescaled 
with the inverse square root of the mean level spacing determined as follows. 
For this value of chemical potential the spreading of data into the complex 
plane is such that an ordering according to the real part does no longer make 
sense.
For a given set of eigenvalues in our window close to the origin 
we thus determine for each eigenvalue its closest neighbor in the complex 
plane and then average over these distances. 
After averaging over all configurations we obtain the 
mean level spacing.
In fig. \ref{2Dcutstr} left (right) the density eq. (\ref{strongrho}) 
is plotted for for fixed $\Re e(\xi)=0$ ($\Im m(\xi)=0$) and compared to the 
histograms along the imaginary (real) axis. We find very good agreement 
with the parameter free prediction\footnote{We note that compared to \cite{AW}
we have refined the determination of the level spacing. In addition 
eq. (\ref{strongrho}) has been rescaled, 
taking properly into account $\qq^2=1/2$.}. 
The statistical fluctuations are still 
larger than in the weak limit, because for about the same number of 
configurations more eigenvalues have moved away from the real axis. 
Along the imaginary axis the density drops to zero already 
at $\sim\pm1.1$ in 
given units, indicating that only part of the left histogram is truly 
in the microscopic regime.

%%%%%%%%%%%%%%%%%%%%%%%%%%%%%%%%%%%%%%%%%%%%%%%%%%%%%%%%%%%%%%%%%%%%%%%%%%%

\section{Conclusions}

To summarize we have shown the equivalence of two different matrix 
model realizations \cite{Steph,A02} 
for QCD with chemical potential. The equivalence
holds for weak non-Hermiticity with several flavors and for strong 
non-Hermiticity with at least one flavor for small chemical potential. 
It shows that the results of \cite{A02} can be directly related to QCD. 
To illustrate this we have compared to quenched QCD lattice data for the 
microscopic Dirac operator spectrum and found agreement in both limits,
for weak and strong non-Hermiticity. 

We close with some remarks on the quenched approximation. While the origin 
of its failure to describe QCD with non-vanishing chemical potential 
was clarified in \cite{Steph}, the matrix model is valid beyond the quenched 
approximation as well. 
It should be mentioned that also for zero chemical potential the quenched 
approximation is not without inconsistencies \cite{Poul}. Nevertheless, 
matrix models do correctly describe unquenched simulations,
as can be seen e.g. for two colors with dynamical fermions \cite{Tilo}.
For complex spectra so far only quenched or phase quenched predictions 
with massless flavors are available \cite{A02}. 
In proving the equivalence we have learned how to introduce mass terms.
This indicates that with a chemical potential the duality 
between massless flavors and topology proved in \cite{ADDV} may no longer 
hold in general. This seems quite plausible since for example the quenched 
theory with nonzero topology $\nu\neq0$ 
should not carry a phase, while the unquenched theory with the 
same number of massless flavors $N_f$ and zero topology should carry a phase.

\indent

{\bf Acknowledgments:} 
I wish to thank Y. Fyodorov, G. Vernizzi and T. Wettig for collaboration
and many fruitful discussions.
This work was supported by a Heisenberg fellowship of the 
Deutsche Forschungsgemeinschaft and in part by the the European network on 
``Discrete Random Geometries'' HPRN-CT-1999-00161 (EUROGRID).

\end{document}